\documentstyle[titlepage,12pt]{article}
\begin{document}
\parindent 2em

\begin{titlepage}
\begin{center}
\vspace{12mm}
{\LARGE Dual superfluid-Bose glass 
critical point in two dimensions and the universal conductivity} 

\vspace{15mm}

Igor F. Herbut\\

Department of Physics and Astronomy, University of British Columbia,\\ 
6224 Agricultural Road, 
Vancouver B. C., Canada V6T 1Z1\\

\end{center}
\vspace{10mm}
{\bf Abstract:}
The continuum version of the dual theory for 
a system of two-dimensional quantum disordered bosons 
with statistical particle-hole symmetry and at $T=0$ 
admits a new disordered 
critical point within the renormalization group calculation 
at fixed dimension. We obtain 
the universal conductivity and the critical exponents at the 
superfluid-Bose glass transition in the system: 
$\sigma_c=0.25 (2e) ^2/h$, 
$\nu=1.38$ and $z=1.93$,  
to the lowest-order  
in fixed-point values of the dual coupling constants. 


\end{titlepage}


The problem of competition between 
superfluidity and localization in  two dimensions (2D) has 
attracted lot of interest, 
as 2D is the lower critical dimension 
for both transitions when considered separately \cite{grinstein}. 
Experiments on 
disordered thin superconducting 
films show a sharp separation 
between samples of varying thickness that exhibit 
diverging or vanishing conductivities as temperature 
is lowered, providing support for the concept 
of quantum ($T=0$) superconductor-insulator (SI) critical point \cite{liu}. 
Under the assumption that the amplitude of the 
superconducting order parameter remains finite at the 
critical value of the tuning parameter, 
it is sensible to describe the SI transition  solely in terms 
of bosonic (phase) degrees of freedom in a  random external potential. 
The resulting problem of disordered interacting bosons 
\cite{halperin}, \cite{weichman} 
is interesting on its own account, and has direct implications for the phase 
transitions in 
other systems like 
$^4 He$ in porous media, Josephson junctions arrays or 
granular superconductors. Not a small part of the fascination with the 
SI transition in 2D comes from a theoretical 
possibility that right at the 
transition point there is a finite dc conductivity,  
and moreover, 
that this conductivity in units of 
$ (2e)^2 /h$ should be a universal number \cite{grinstein}. The  
universal metallic behavior at the SI transition is an option 
unique for 2D, which posed a challenge for analytical calculations 
of the critical conductivity \cite{cha}, 
particularly in the disordered case where 
this quantity has 
hitherto been obtained only 
numerically \cite{batrouni}. Generally, the critical behavior 
of disordered interacting bosons in more than one dimension 
\cite{giamarchi} has proven very difficult to address analytically
\cite{kim}, \cite{zhang}, partly due to 
the fact that there is no easily identifiable  
correct mean-field starting point around which to set up a 
systematic perturbative calculation. 

 In this Letter we present the results of a new approach to 
the quantum critical behavior of 
2D disordered bosonic systems, applied to the 
paradigmatic problem of disordered lattice bosons 
at a commensurate density, 
interacting via short-range repulsion. Without disorder and at 
$T=0$ the system (described by the 
Hamiltonian (1)) undergoes a Mott insulator-superfluid 
(MI-SF) transition as the ratio $t/U$ is varied, 
which is due to a commensurable boson density 
in the universality class of the 3D XY model \cite{weichman}. 
With disorder the transition  into a superfluid state is expected to 
always proceed from the gapless Bose glass (BG) phase \cite{weichman}. 
We study the SI transition in this system within the continuum 
version of the {\it dual} theory for the bosonic 
Hamiltonian \cite{lee}, which describes topological defects in the 
quantum ground state, and for the problem at hand takes 
the form of 3D classical anisotropic 
scalar electrodynamics (Ginzburg-Landau 
superconductor) in random magnetic field. 
The dual formulation of the problem 
may be thought of 
as a higher-dimensional 
analog of the Haldane's density representation of dirty bosons in 1D
\cite{giamarchi}. Since in dual representation the 
Gaussian point represents the {\it interacting} superfluid phase 
of the original bosons, one could expect that perturbing around it 
may be more sensible than in the original formulation, where one expands 
around the limit of non-interacting particles. Critical 
behavior of the isotropic version of the dual 
theory without disorder has been studied recently in different context by 
perturbative and non-perturbative methods \cite{parisi}, 
\cite{tesa}, \cite{berg}. 
While a small anisotropy at the critical point of the pure theory 
is {\it irrelevant}, 
the correlated random magnetic field 
is exactly {\it marginal} perturbation to the lowest order in disorder. 
However, we show that it generates 
a correlated random-mass term  for the disorder field, 
which represents 
a strongly {\it relevant}  
perturbation at the pure critical point. 
In one-loop approximation and 
for the physical initial values of the coupling constants the flow 
at the critical surface is 
attracted to a new disordered fixed point, which 
governs the critical behavior at the 
BG-SF transition in the system. 
We calculate the universal {\it resistivity} at the BG-SF transition,  
as well as the correlation length and the dynamical 
critical exponents $\nu$ and $z$, as series in 
fixed-point values of the dual coupling constants, to the lowest 
non-trivial order.  
The comparison with previous studies and 
with experiments is discussed. 

  Specifically, we are interested in the 
Hamiltonian for a 
collection of charge-2e bosons with short-range repulsion, $U>0$, 
written in the number-phase representation
\cite{weichman}, \cite{lee}: 
\begin{equation}
\hat{H} = U \sum_{i} \hat{n}_{i}^2 -\sum_{i} h_i \hat{n}_i 
-t\sum_{\langle i,j\rangle} cos( \hat{\phi}_i - \hat{\phi}_j ), 
\end{equation}
 where index $i$ labels the sites of a 2D lattice, $\hat{n}_i$ represents 
the deviations from a large 
{\it integer} average number of bosons per site, 
$\{h_i\}$ are Gaussian random variables with zero mean and width 
$w_h$, and $\hat{\phi}_i$ 
is the phase variable canonically 
conjugate to the number of bosons, $[\hat{\phi}_i, 
\hat{n}_{j}]=i\delta_{ij}$. The Josephson term couples 
only the nearest neighbours. 
Using a  
set of exact and nearly-exact transformations on a lattice 
\cite{dasgupta}, \cite{herbut} 
Fisher and Lee \cite{lee} 
have demonstrated that the partition function for the disordered 
problem at $T=0$ may be written in the form of 
classical 3D (two spatial and one 
imaginary time dimensions) strongly anisotropic
lattice superconductor in a random magnetic field. The dual 
lattice theory should be in the same universality class as the 
continuum field theory for a strongly 
type-II superconductor ($\lambda >> q^2$)  \cite{herbut}: 
\begin{eqnarray}
S=\int d^2 \vec{r} dz [ |(\nabla - i q \vec{A}(\vec{r},z))\Psi(\vec{r},z)
|^2 + \mu^2 |\Psi(\vec{r},z)|^2 + \frac{\lambda}{2}|\Psi(\vec{r},z)|^4 
\nonumber \\
+ \frac{1}{2} (\nabla \times A(\vec{r},z))^2 
+ \frac{\alpha}{2} 
(\nabla \times \vec{A}(\vec{r},z))^2 _{\hat{z}}+ 
h(\vec{r}) (\nabla \times \vec{A}
(\vec{r},z)) _{\hat{z}}], 
\end{eqnarray}
where the complex $\Psi$ represents the
 disorder-field \cite{kleinert}. The   
condensation of $\Psi$ implies the proliferation of vortices in the 
ground state and consequently, the {\it destruction} of superfluidity 
by giving (Higgs) mass to the superfluids 
sound mode,  
represented by the dual gauge-field $\vec{A}$ 
($\nabla\cdot\vec{A}=0$). Note 
that the random magnetic field along $\hat{z}$ (time) 
direction $h(\vec{r})$ depends only on two 
(real space) coordinates, so that 
disorder is correlated along the 
third direction, as typical for quantum problems with quenched disorder. 
The condition on the average field $\overline{h}=0$ corresponds to 
integer average number of particles per site. 
 
The SI transition in dual theory is reached 
by tuning the renormalized mass of the disorder-field to zero. Introducing 
replicas in standard way to average over disorder, 
the dual theory at $m =0$ may be 
parametrized by four dimensionless coupling constants: the dual charge 
$\hat{q}^2=q^2 /p$, the quartic term coupling 
$\hat{\lambda}=\lambda/p$, the anisotropy parameter 
$\alpha$, and the coupling $\hat{w}_h=
w_h/p$, which appears in the disorder-induced term of the form 
$-(w_h/2)\sum_{\alpha, \beta}\int (\nabla\times\vec{A}_{\alpha}(\vec{r},z))
_{\hat{z}}
(\nabla\times\vec{A}_{\beta}(\vec{r},z'))_{\hat{z}}$.  
$p$ is an arbitrary infrared scale. 
Besides these, the random magnetic field 
{\it generates} yet another quartic coupling constant 
$\hat{w}=w/\pi p^2$ via the  term $-(w/2)\sum_{\alpha,\beta}\int 
|\Psi_{\alpha}(\vec{r},z)|^2 |\Psi_{\beta}(\vec{r},z')|^2$ 
in the replicated theory, with   
a meaning of width 
of the Gaussian distribution of the random-mass term (again correlated 
along $\hat{z}$ axis) for the disorder-field. 
To determine the flow of the five coupling constants 
we performed a perturbative RG calculation in fixed dimension 
\cite{parisi}, \cite{tesa}, 
\cite{berg}, 
since we will eventually be interested in the 
critical conductivity, which is finite only in 2D. 
The gist of the method is to express the universal quantities as series in 
renormalized, instead of bare, coupling constants, so they tend to 
finite values in the infrared limit, determined only by the location of 
the attractive fixed point. We define the renormalized 
couplings $\lambda$ and $w$ as the corresponding quartic vertices 
right at $m=0$ and at the usual symmetric point 
in the momentum space: 
$ \vec{k}_i \cdot \vec{k}_{j} = (4\delta_{i,j} -1)p^2 /4$, $i,j=1,2,3$,  
with additional condition $k_{1,z}=k_{3,z}=-k_{2,z}=-k_{4,z}$ to 
accommodate the correlated nature of the disorder vertex $w$ along $z$-axis.
The polarization diagram which renormalizes $q$, $\alpha$ 
and $w_h$ is evaluated at the external momentum 
$c\cdot p$, 
where the parameter $c$ will be specified shortly.
The $\beta$-functions 
for the anisotropy $\alpha$ and the random-field 
width  $w_{h}$ are:
\begin{equation}
\frac{d\alpha}{dt}= -\alpha (\frac{c}{16}\hat{q}^2+ O(\hat{q}^4))
+O(\alpha^2),
\end{equation} 
\begin{equation}
\frac{d\hat{w}_{h}}{dt} = \hat{w}_{h} (1-\eta_{A}) + O(\hat{w}_{h}^{2}), 
\end{equation}
where $t=\ln(1/p)\rightarrow \infty$, in the infrared limit, and 
renormalized couplings are assumed. 
Eq. 3  
is a one-loop result, and suggests that a small anisotropy is 
{\it irrelevant} at any fixed point with a finite dual 
charge. Although the dual theory 
should be strongly and not weakly anisotropic \cite{lee}, 
we do not expect 
stable fixed points 
with $\alpha\neq 0$. This is known to be true in the 
classical scalar electrodynamics close to four dimensions 
and for a large number of complex field  components 
\cite{lubenski}. In contrast to eq. 3, eq. 4 is {\it exact} 
to the lowest order in $\hat{w}_{h}$. This follows from the fact 
that the transverse part of 
polarization is diagonal in replica indices to the 
lowest order in $\hat{w}_{h}$, so that the renormalization 
of $\hat{w}_{h}$ to this order 
entirely comes from rescaling of the gauge-field. 
The anomalous dimension of the 
gauge-field propagator is $\eta_A=1$ in 3D at any 
fixed point with 
a finite dual charge 
\cite{tesa}. In particular, the linear term in 
eq. 4 vanishes at the pure MI-SF fixed point (see below). This 
observation may explain why numerical calculations at 
commensurate boson densities always have difficulties in 
detecting the intervening Bose glass phase at weak disorder 
\cite{kisker}. It also seems 
consistent with the result of Singh and Rokshar \cite{zhang}, where a weak 
disorder in bosonic Hubbard 
model is found to be irrelevant, 
but only unusually weakly so. 
Fate of the coupling constant $\hat{w}_h$ is determined by the 
higher order terms in eq. 4.  To proceed, we 
will {\it assume} that $\hat{w}_h$ is irrelevant at the MI-SF fixed 
point of the theory \cite{remark}, and that its only role 
was to generate a strongly relevant coupling $\hat{w}$, before 
renormalizing to zero.
Hereafter we will therefore set both $\alpha=\hat{w}_h =0$. 

 One-loop $\beta$-functions for the remaining three coupling 
constants are then:
\begin{equation}
\frac{d\hat{q}^2}{dt}= \hat{q}^2 - \frac{c}{16} \hat{q}^{4},
\end{equation}
\begin{equation}
\frac{d\hat{\lambda}}{dt}=\hat{\lambda}(1+\frac{1}{2}\hat{q}^2 +
2 ( \sqrt{2} \ln(1+\sqrt{2}) +\frac{1}{\sqrt{6}} \ln( \sqrt{2}+
\sqrt{3}) ) \hat{w}) -\frac{2\sqrt{2}+1}
{8} \hat{\lambda}^{2}-\frac{1}{2\sqrt{2}} \hat{q}^4, 
\end{equation}
\begin{equation}
\frac{d\hat{w}}{dt}= \hat{w} (2+\frac{1}{2}\hat{q}^{2}-\frac{1}{2\sqrt{2}}
\hat{\lambda}) + 2( \frac{1}{\sqrt{2}} \ln(1+\sqrt{2}) + \frac{1}{\sqrt{6}} 
\ln(\sqrt{2}+\sqrt{3}) )\hat{w}^{2}.  
\end{equation}
For $\hat{w}=0$ these equations reduce to those studied previously 
in the context of critical behavior of superconductors \cite{tesa}. 
In that case for a choice $c>5.17$ \cite{note} 
there are four fixed points 
in the theory: Gaussian ($\hat{q}^2=\hat{\lambda}=0$) and 3D XY 
($\hat{q}^2=0$, $\hat{\lambda}_{xy}=2.09$), both unstable in direction 
of the dual charge, tricritical ($\hat{q}^2=16/c$, $\hat{\lambda}=
\hat{\lambda}_{-}$), unstable in $\hat{\lambda}$-direction, and the 
MI-SF critical point ($\hat{q}^{2}=16/c$, $\hat{\lambda}=
\hat{\lambda}_{+}$), 
believed to be of "inverted" XY type \cite{dasgupta} 
(see Figure 2. in ref. 12). 
The Gaussian and the tricritical fixed points  
are connected by the straight separatrix 
determined by the tricritical value of the Ginzburg-Landau 
parameter $\kappa=\sqrt{\lambda /2q^2}=\kappa_c$ in the problem.
On the other hand, the tricritical value of $\kappa$ has 
been estimated both analytically \cite{berg} 
and numerically \cite{bartho}, which allows us to 
fix the parameter $c$ to 
match that number: $\kappa_c=0.42/\sqrt{2}$ requires $c=20$. 
It is worth noting that for $\hat{w}=0$ 
our calculation gives reasonable estimates of the 
critical exponents both at the unstable XY ($\nu=0.63$) and 
at the stable MI-SF fixed point ($\nu=0.61$, $\eta=-0.20$). 

A small $\hat{w}$ is relevant at the MI-SF fixed point, as  
follows from eq. 7  and from the 
Harris criterion \cite{chayes}. Both 
perturbative \cite{tesa} and non-perturbative \cite{berg}, \cite{dasgupta}, 
\cite{herbut} calculations of the correlation 
length exponent at the pure critical point yield 
values $\nu<1$, suggesting 
relevance of the correlated random-mass term. 
Starting from the type-II region ($\lambda>>q^2$), the flows are  
attracted by the disordered critical point at 
$\hat{q}^{2}_c=0.8$, $\hat{w}_c = 3.71  $ and $\hat{\lambda}_c = 29.72 $, 
which we identify as the BG-SF critical point. 
The BG-SF critical point exists for any choice of renormalization procedure 
(parameterized by $c$), unlike the charged 
fixed points in the pure theory which exist only for $c>5.17$. 
This is akin to the 
classical scalar electrodynamics close to 4D \cite{cardy}, where 
randomness restores the critical point in  the theory. 
To the lowest order in the critical 
coupling constants the exponents at the BG-SF fixed point are:  
\begin{equation}
\nu=\frac{1}{2} (1+\frac{\hat{\lambda}_c - \hat{q}^2 _c-4\hat{w}_c}{8})=1.38,
\end{equation}
\begin{equation}
z=1+\frac{\hat{w}_c}{4}= 1.93 .   
\end{equation}
Note that $\nu>1$, as expected \cite{chayes}. 
The dynamical exponent is very close 
to two, which was conjectured to be an exact result \cite{weichman} in 2D. 
The exponents are also in 
good agreement with Monte 
Carlo calculations of Wallin et al. \cite{batrouni} 
($\nu=0.9\pm 0.1$, $z=2\pm 0.1$) and real-space study of Zhang and Ma 
\cite{zhang} ($\nu=1.4$, $z=1.7$). 

The disorder-field 
$\Psi$ describes condensation of topological defects in the 
bosonic ground state, and duality of charges and vortices implies 
that the dimensionless conductance for vortices is the {\it resistance} 
for the original bosons. Kubo formula for the 
conductivity (or conductance, in 2D) of the vortices described 
by the dual theory
then leads to the critical dc resistivity at $T=0$:
\begin{equation}
R_c =\frac{2\pi h}{(2e)^2} 
\lim_{\omega\rightarrow 0} \frac{1}{\omega}
[ 2\langle \overline{\Psi^{*}(0) \Psi(0)\rangle} -
\int d^2 \vec{r} dz e^{i\omega z} \langle \overline { j_{x}(\vec{r},z)  
j_{x}(\vec{0},0)} \rangle],   
\end{equation}
where $\vec{j}$ is the current of the dual field. 
To the lowest order 
we obtain
\begin{equation}
R_c = (\frac{\pi}{8}+\frac{\pi}{16}\hat{q}^{2}_c+
 \frac{2.94}{\pi}\hat{w}_c)
\frac{h}{(2e)^2}. 
\end{equation}
The first term describes the  Gaussian part of the 
resistivity at the critical point \cite{cha}, while the  
second term accounts for the 
lowest order wave-function renormalization.  
The third, disorder term, is the main source 
of dissipation. 
Inserting the values of the critical 
coupling constants into eq. 11 we estimate the 
universal conductivity for the bosons:
\begin{equation}
\sigma_c=R_c  ^{-1} = 0.25 \frac{(2e)^2}{h}. 
\end{equation} 
The result is somewhat larger than in 
Monte Carlo calculations of Wallin et al. \cite{batrouni}: 
$\sigma_c =[0.14\pm 0.03](2e)^2 /h$, and smaller than in Batrouni et al. 
\cite{batrouni}: $\sigma_c =[0.45\pm 0.07](2e)^2 /h$. 
Our preliminary calculations 
suggest that the higher order 
corrections tend to lower the value of $\sigma_c$. 
Remark however that the value of $\sigma_c$ is already
 smaller than in the pure case 
\cite{cha}. 

  The reader has certainly noted 
that the BG-SF fixed point in question 
is a strong-coupling one, and the numerical values of critical 
quantities cited above need to 
be taken with some reservation. There is no small parameter in the 
problem and we had to rely on a less controlled calculational scheme  
in fixed dimension \cite{parisi}. 
Also, a different tricritical value 
of GL parameter $\kappa_c$ 
in zero-disorder scalar electrodynamics would dictate a 
different choice of the parameter $c$,  
and would 
affect the position of the SF-BG fixed point within our 
calculation. The dependence 
of the critical quantities on this 
variation seems weak, fortunately. 
For instance, assuming almost a twice larger 
value for $\kappa_c=0.8/\sqrt{2}$ \cite{kleinert} 
we would find $\nu=1.55$, $z=2.14$ 
and $\sigma_c=0.18(2e)^2 /h$.
Our calculation allows for a systematic improvement by 
going to higher orders in the dual coupling constants and using 
standard resummation techniques \cite{zinn}. 
Although the perturbation 
series in eqs. 3-7 are likely to be only asymptotic (but Borel 
summable \cite{zinn}), 
it is encouraging 
that the lowest order calculation already produces sensible results.  

We have observed that our numbers for the critical quantities
are not too different from those obtained by other methods 
 performed on the Hamiltonian (1) \cite{batrouni} or on its 
variant \cite{zhang}, but at  
a {\it non-commensurate} density of bosons. 
Without disorder, at non-commensurate filling factors 
the MI-SF transition is in a different universality 
class, and it is in fact mean-field in character \cite{weichman}. 
One would expect on physical 
grounds that commensurability may 
be irrelevant at the disordered critical point \cite{kim}.  
Deviation from a commensurate density would correspond to a finite 
average magnetic field $\overline{h}$ in dual theory, 
whose scaling dimension 
at the BG-SF fixed point we do not presently know. 
We have argued however, that even at  commensurate densities the 
transition is always controlled by the disordered BG-SF fixed point, 
due to the generated random-mass term for the disorder 
field. 

Comparison with experiments is made difficult by the fact that 
different measurements lead to rather different values of 
the three critical quantities we considered \cite{liu}. We may still note 
that our numbers for the critical exponents fall 
within the range $\nu z=2.8\pm 0.4$ 
obtained from the scaling of resistivity 
on the insulating side of the SI transition in thin Bi films 
\cite{liu}. The value of the critical conductivity in thin films 
differs by a factor 
of $\sim 4$ in various experiments \cite{liu}, 
and its universality may be questioned. 
 Many experiments show $\sigma_c \sim  (2e)^2 /h$ \cite{liu}, 
somewhat larger than our estimate. Note also 
that experimentally one typically measures  $\omega=0$, 
$T\rightarrow 0$ conductivity \cite{giamarchi}, 
which is the opposite order of limits 
than assumed here. It has been argued recently that 
the two limits may not commute \cite{damle}, and that the 
critical conductivity has a non-trivial structure as a 
function of $\omega/T$.  	

In conclusion, we considered the critical behavior 
at the superfluid-insulator transition 
in a system of 2D disordered bosons with commensurate density 
and at $T=0$, by studying the continuum 
version of the dual theory that describes vortices in the 
superfluid ground state. Although disorder, which 
initially appears in the dual theory as  
random magnetic field, is argued to be marginally irrelevant  at 
the critical point of the pure theory, it generates another strongly 
relevant disorder-like coupling constant,  and the transition 
is ultimately governed by the  BG-SF critical point. 
The universal resistivity at the BG-SF transition 
and the critical exponents have been expressed 
as perturbation series in the critical coupling constants, and the lowest 
order results 
compare well with the 
 Monte Carlo calculations performed on the original 
boson Hamiltonian. 

 The author is thankful to Professors 
I. Affleck, P. Stamp and Z. Te\v sanovi\' c for 
useful discussions. This work has been supported by NSERC of Canada 
and Izaak Walton Killam foundation.



\begin{thebibliography}{99}
\bibitem{remark} We may formally assure this by staying infinitesimally 
(by $\epsilon$) below one time-dimension in the dual theory,
so that 
the scaling dimension of $\hat{w}_h$ at the MI-SF fixed point 
becomes $-2\epsilon$. At $\epsilon=0$ 
however, $\hat{w}_h$ can at best be marginally irrelevant.
\bibitem{note} In RG in fixed dimension the coefficients of perturbative 
$\beta$-functions in eq. 3-7 are procedure dependent, as 
is reflected here in the appearance 
of the parameter $c$ in eqs. 3 and 5. This suggests a 
choice of $c$ based on prior knowledge of some feature of 
the RG flows, as discussed next 
in the text. See also ref. 12.
\end{thebibliography}
\end{document}